\documentclass[%
 reprint,
 superscriptaddress,
 groupedaddress,
 preprintnumbers,
 bibnotes,
 amsmath,amssymb,
 aps,
longbibliography
]{revtex4-1}

 \usepackage{todonotes}

\usepackage[utf8]{inputenc}
\usepackage{graphicx}
\usetikzlibrary{shapes.misc}
\usetikzlibrary{cd}
\usepackage{dcolumn}
\usepackage{bm}
\usepackage{hyperref}
\hypersetup{
    colorlinks=true,
    linkcolor=black,
    filecolor=black,      
    urlcolor=black,
    citecolor=black,
    breaklinks
}

\newcommand{\me}{\mathrm{e}}

 \newcommand{\Ef}[3]{{\textrm{E}_4}\!(\begin{smallmatrix}#1\\#2\end{smallmatrix};#3)} 
 \newcommand{\gamt}[3]{{\widetilde{\Gamma}}\!(\begin{smallmatrix}#1\\#2\end{smallmatrix};#3)} 

\begin{document}

\title{The elliptic double box and symbology beyond polylogarithms}
\author{Alexander Kristensson}
\author{Matthias Wilhelm}
\author{Chi Zhang}
 \affiliation{Niels Bohr International Academy, Niels Bohr Institute, Copenhagen University, Blegdamsvej 17, 2100 Copenhagen \O{}, Denmark}

\begin{abstract}
We study the elliptic double-box integral, which contributes to generic massless QFTs and is the only contribution to a particular 10-point scattering amplitude in $\mathcal{N}=4$ SYM theory. Based on a Feynman parametrization, we express this integral in terms of elliptic polylogarithms. We then study its symbol, finding a rich structure and remarkable similarity with the non-elliptic case. In particular, the first entry of the symbol is expressible in terms of logarithms of dual-conformal cross-ratios, and elliptic letters only occur in the last two entries. Moreover, the symbol makes manifest a differential equation relating the double-box integral to a 6D hexagon integral, suggesting that it can be bootstrapped based on the latter integral alone. 
\end{abstract}

\maketitle

\section{Introduction}

Understanding of numbers and functions in QFT in general and in $\mathcal{N}=4$ SYM theory in particular has lead to great progress in calculating scattering amplitudes as well as other quantities. 

For one-loop quantities, multiple polylogarithms (MPLs) ~\cite{Chen:1977oja,G91b,Goncharov:1998kja,Remiddi:1999ew,Borwein:1999js,Moch:2001zr} suffice, and this continues to be the case in massless theories for higher loop orders for sufficiently low numbers of external particles.
MPLs are characterized via their symbol \cite{Goncharov:2010jf},
a tensor product -- or word -- in so-called letters $\log(\phi_\alpha)$, where $\phi_\alpha$ are functions of the kinematic variables. These letters encode the singularity and branch-cut structure and their union is known as symbol alphabet. 
In cases where the $\phi_\alpha$ are rational (or can be simultaneously rationalized), the symbol has made it possible to bootstrap the corresponding amplitudes to very high loop orders, i.e.\ to make an ansatz based on an assumed symbol alphabet \cite{Golden:2013xva} and to fix the coefficients via various constraints such as the Steinmann conditions \cite{Steinmann,Steinmann2,Caron-Huot:2019bsq} and cluster adjacency \cite{Drummond:2017ssj,Drummond:2018dfd}; see e.g.\  \cite{Caron-Huot:2020bkp,Caron-Huot:2019vjl,Drummond:2018caf,Dixon:2016nkn,Dixon:2020cnr,Dixon:2020bbt}.
For slightly more legs, however, also symbol alphabets with $\phi_\alpha$ occur that are not simultaneously rationalizable \cite{Bourjaily:2019igt,Zhang:2019vnm,He:2020vob,He:2020lcu}.

Beyond MPLs, infinite towers of more complicated functions occur \cite{Brown:2009ta,Brown:2010bw,Bourjaily:2018yfy,Bourjaily:2018ycu,Festi:2018qip,Broedel:2019kmn,Besier:2019hqd,Bourjaily:2019hmc,Vergu:2020uur,mirrors_and_sunsets,
Broadhurst:1987ei,Adams:2018kez,Adams:2018bsn,Huang:2013kh}.
The simplest of these classes of functions involve integrals over elliptic curves; they have recently been increasingly well understood in terms of so-called elliptic multiple polylogarithms (eMPLs)
\cite{Laporta:2004rb,MullerStach:2012az,brown2011multiple,Bloch:2013tra,Adams:2013nia,Adams:2014vja,Adams:2015gva,Adams:2015ydq,Adams:2016xah,Adams:2017ejb,Adams:2017tga,Bogner:2017vim,Broedel:2017kkb,Broedel:2017siw,Adams:2018yfj,Broedel:2018iwv,Broedel:2018qkq,Honemann:2018mrb,Bogner:2019lfa,Broedel:2019hyg,Duhr:2019rrs,Walden:2020odh,Weinzierl:2020fyx}.
In particular, also a symbol for eMPLs has been defined \cite{BrownNotes,Broedel:2018iwv}; the symbol letters in this case are $\Omega^{(j)}(\tilde\phi_\alpha)$, where $\tilde\phi_\alpha$ are functions of the images of the kinematics when mapped to the torus, which is equivalent to the elliptic curve.

\begin{figure}
\centering
\begin{tikzpicture}[label distance=-1mm]
		\node[label=left:$8$] (0) at (0, 15.5) {};
		\node[label=above:$10$] (1) at (1.5, 16) {};
		\node[label=above:$9$] (2) at (0.5, 16) {};
		\node[label=above:$1$] (3) at (2.5, 16) {};
		\node[label=right:$2$] (4) at (3.0, 15.5) {};
		\node[label=below:$6$] (5) at (0.5, 14) {};
		\node[label=left:$7$] (6) at (0, 14.5) {};
		\node[label=below:$5$] (7) at (1.5, 14) {};
		\node[label=right:$3$] (8) at (3, 14.5) {};
		\node[label=below:$4$] (9) at (2.5, 14) {};
        \node(11) at (5, 15) {};
		\node (12) at (6, 15) {};
		\draw[thick] (0.center) to (4.center);
		\draw[thick] (6.center) to (8.center);
		\draw[thick] (2.center) to (5.center);
		\draw[thick] (1.center) to (7.center);
		\draw[thick] (3.center) to (9.center);
        \node[label=left:\textcolor{blue!50}{$x_8$}] (10) at (0.2, 15) {};
        \node[label=above:\textcolor{blue!50}{$x_{10}$}] (13) at (1.0, 15.8) {};
        \node[label=above:\textcolor{blue!50}{$x_1$}] (14) at (2.0, 15.8) {};
        \node[label=right:\textcolor{blue!50}{$x_3$}] (15) at (2.8, 15) {};
        \node[label=below:\textcolor{blue!50}{$x_5$}] (16) at (2.0, 14.3) {};
        \node[label=below:\textcolor{blue!50}{$x_6$}] (17) at (1.0, 14.3) {};
		 \draw[very thick,blue!50] (10.center) to (15.center);
		 \draw[very thick,blue!50] (13.center) to (17.center);
		 \draw[very thick,blue!50] (14.center) to (16.center);
       \draw [thick] (4.9,15.4) -- (4.9,14.6);
       \draw [thick] (6.2,15.4) -- (6.2,14.6);
       \draw [thick] (6.2,15.4) -- (5.55,15.8) -- (4.9,15.4);
       \draw [thick] (4.9,14.6)-- (5.55,14.2) --  (6.2,14.6);
       \draw [thick] (5.55,15.8) --(5.55,16.1) ;
       \draw [thick] (5.55,14.2)-- (5.55,13.9);
       \draw [thick] (4.9,15.4) -- (4.6, 15.7);
       \draw [thick] (4.9,15.4) -- (4.5, 15.5);
       \draw [thick] (4.9,14.6) -- (4.6, 14.3);
       \draw [thick] (4.9,14.6) -- (4.5, 14.5); 
       \draw [thick] (6.2,15.4) -- (6.5,15.7);
       \draw [thick] (6.2,15.4) -- (6.6,15.5);
       \draw [thick] (6.2,14.6) -- (6.5,14.3);
       \draw [thick] (6.2,14.6) -- (6.6,14.5); 
       \node at (5.55,16.3) {10};
       \node at (5.55,13.7) {5}; 
       \node at  (6.60,15.85) {1};
       \node at  (6.75,15.45) {2};
       \node at  (6.75,14.55) {3};
       \node at  (6.60,14.15) {4};
       \node at  (4.50,14.15) {6};
       \node at  (4.35,14.55) {7};
       \node at  (4.35,15.45) {8};
       \node at  (4.50,15.85) {9};
\draw[very thick,blue!50] (5.55,15) -- (5.1,15.8) ;
\draw[very thick,blue!50] (5.55,15) -- (6.0,15.8) ;
\draw[very thick,blue!50] (5.55,15) -- (4.7,15) ;
\draw[very thick,blue!50] (5.55,15) -- (6.4,15) ;
\draw[very thick,blue!50] (5.55,15) -- (5.1,14.2) ;
\draw[very thick,blue!50] (5.55,15) -- (6.0,14.2) ;
\node at  (5.1,15.95) {\textcolor{blue!50}{$x_{10}$}};
\node at  (6.05,15.95) {\textcolor{blue!50}{$x_{1}$}};
\node at (6.65,15) {\textcolor{blue!50}{$x_{3}$}};
\node at (4.50,15) {\textcolor{blue!50}{$x_{8}$}};
\node at (6.05,14) {\textcolor{blue!50}{$x_{5}$}};
\node at (5.1,14) {\textcolor{blue!50}{$x_{6}$}};
\end{tikzpicture} 
\caption{The elliptic double box and the related 6D hexagon, as well as their dual graphs.
}
\label{fig: integral}
\end{figure}
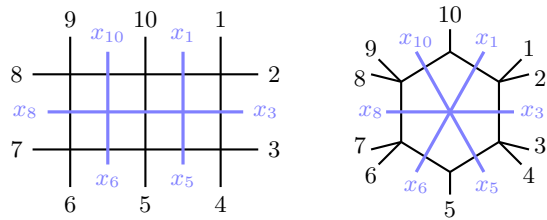

In $\mathcal{N}=4$ SYM theory, the first time elliptic functions occur is the 10-point N${}^3$MHV amplitude at two-loop order. A particular component of it is given in terms of a single Feynman diagram, the elliptic double-box integral \cite{CaronHuot:2012ab}, depicted in fig.\ \ref{fig: integral}.
This integral was found to satisfy a first-order differential equation relating it to the 6D hexagon \cite{Paulos:2012nu,Nandan:2013ip}, as well as a further second-order differential equation \cite{Chicherin:2017bxc}.
A four-fold rational integral representation -- and a one-fold polylogarithmic one -- were found \cite{Bourjaily:2017bsb}, as well as a sum representation \cite{Loebbert:2019vcj,Ananthanarayan:2020ncn}.
So far, however, it has not been possible to express the elliptic double-box integral in terms of eMPLs.

In this paper, we express the double-box integral in terms of eMPLs and calculate its symbol, finding a rich structure. 
In particular, we observe that the symbol satisfies the first-entry condition occurring for MPLs \cite{Gaiotto:2011dt}: the letters $\Omega^{(j)}(\tilde\phi_\alpha)$ in the first entry combine to $\log(u)$, where $u$ is a dual conformal cross-ratio. Similarly, the letters in the second entry combine to $\log$s, such that elliptic letters only occur in the last two entries. 
Moreover, the symbol makes manifest the differential equation relating the elliptic double-box integral to the  6D one-loop hexagon integral. 

The remainder of this paper is structured as follows. 
In section \ref{sec: linear reducibility}, we solve the linear-reducibility problem that has so far prevented integration in terms of eMPLs. 
In section \ref{sec: to the torus}, we transition from the description in terms of the elliptic curve to one in terms of the torus, which more naturally leads to the symbol.
We discuss the properties of the symbol 
in section \ref{sec: symbol}.
Our conclusion and outlook is contained in section \ref{sec: conclusion}.

\section{The linear reducibility problem in the double box and its resolution}
\label{sec: linear reducibility}

Let us start with the dual conformal Feynman parameter representation of the elliptic double box \cite{Bourjaily:2017bsb}:
\begin{equation} \label{feyofdb}
    I^{\text{ell}}_{\text{db}}=\int_{0}^{\infty}d ^{4}\vec{\beta}\,\frac{
    1
    }{f_{1}f_{2}f_{3}}\,,
\end{equation}
where 
\begin{align}
    &f_{1}=\beta_{4}(1{+}\beta_{1}){+}\beta_{1}\:,\quad 
    f_{2}=1{+}u_{1}\beta_{4}{+}v_{1}\beta_{1}{+}u_{2}\beta_{2}{+}v_{2}\beta_{3},  \nonumber \\
    &f_{3}=(1{+}u_{3}\beta_{4})\beta_{2}{+}(1{+}u_{4}\beta_{1})\beta_{3}{+}\beta_{2}\beta_{3}{+}u_{3}u_{4}u_{5}f_{1}. \label{defoff}
\end{align}
The cross-ratios are defined by
\begin{gather}
    u_{1}=x_{1,3;5,8} \,,\: 
    u_{2}=x_{3,6;8,10}\,,\:
    v_{1}=x_{1,8;5,3} \,,\:
    v_{2}=x_{3,10;8,6}\,,\nonumber \\
    u_{3}=x_{1,3;5,10} \,,\quad 
     u_{4}=x_{1,6;5,3} \,,\quad 
    u_{5}=x_{1,5;6,10} \,, \label{defuv} 
\end{gather}
where
\begin{equation}
 x_{ab;cd}=\frac{x_{ab}^{2}x_{cd}^{2}}{x_{ac}^{2}x_{bd}^{2}}
\end{equation}
with $x_{ab}=x_a-x_b$ and dual momenta defined as $x_{a}-x_{a+1}=p_{a}$.

In addition to the manifest dual conformal symmetry, the double-box integral has two reflections symmetries $R_1$ and $R_2$ along the horizontal and vertical direction in fig.\ \ref{fig: integral}. The action of these symmetries on the cross-ratios is discussed in the supplementary material. 

As indicated in~\cite{Bourjaily:2017bsb}, three integrations can be performed in terms of polylogarithms~\footnote{The integration order adopted here is different from \cite{Bourjaily:2017bsb}: we integrate out $\beta_{3}$ and $\beta_{4}$ first, then $\beta_{2}$.}, such that the double-box integral can schematically be expressed as 
\begin{equation}
    I^{\text{ell}}_{\text{db}}\sim\int_{0}^{\infty} \frac{d  \beta_{1}}{\sqrt{Q(\beta_{1})}} H(\beta_{1})\,,
\end{equation}
where $Q(\beta_{1})$ is an irreducible quartic polynomial in $\beta_{1}$ and $H(\beta_{1})$ is a pure combination of MPLs of weight three. The obstacle in performing the last integration in terms of elliptic polylogarithms is that the letters of $H(\beta_{1})$ involve not only $\sqrt{Q(\beta_{1})}$ but also square roots of the two quadratic polynomials in $\beta_{1}$. These polynomials share no roots, hence there is no way to rationalize the square roots of two quadratics without increasing the degree of $Q(\beta_{1})$.

To overcome this obstacle, one needs to trace the origin of these additional square roots which are related to the linear reducibility problem of the Feynman parameter integrals \cite{Brown:2009ta,Panzer:2014gra,Broedel:2019hyg,Bourjaily:2021lnz}. In our case, these square roots of quadratics are introduced in the third integration. More precisely, consider the integral after integrating out $\beta_{3}$ and $\beta_{4}$,
\begin{equation}
    \int_{0}^{\infty}d ^{4}\vec{\beta}\,\frac{
    1
    }{f_{1}f_{2}f_{3}}=\int_{0}^{\infty} \frac{d \beta_{1}d \beta_{2}}{\mathcal{P}(\beta_{1},\beta_{2})} \mathcal{G}_{2}(\beta_{1},\beta_{2})\,,
\end{equation}
where the polynomial $\mathcal{P}$ has degree 3 and 2 in $\beta_{1}$ and $\beta_{2}$ respectively, and $\mathcal{G}_{2}(\beta_{1},\beta_{2})$ is a pure combination of MPLs of weight two. Three of the letters of $\mathcal{G}_{2}(\beta_{1},\beta_{2})$ are \emph{quadratic} in $\beta_{1}$ and $\beta_{2}$, while the other letters are linear in $\beta_{1}$ and $\beta_{2}$. It is these three letters that introduce additional square roots in the third integration. To perform the third integration without introducing additional square roots, one needs to make a variable substitution for $\beta_{1},\beta_{2}$ such that the letters of $\mathcal{G}_{2}$ are linear in one of the new integration variables. A crucial observation here is the following: these three letters, which we denote by $q_{1},q_{2}$, and $q_{3}$, can be expressed as
\begin{align}
    q_{1}&=\beta_{1}(\beta_{2}u_{2}+\beta_{1}v_{1})+\cdots \:, \nonumber \\
    q_{2}&=-u_{3}(\beta_{2}+\beta_{1}u_{4}u_{5})(\beta_{2}u_{2}+\beta_{1}v_{1})+\cdots \:,  \\
    q_{3}&=(\beta_{2}+\beta_{1}u_{4}u_{5})(\beta_{2}u_{2}+\beta_{1}v_{1})+\cdots \:, \nonumber 
\end{align} 
where ``$\cdots$'' denote terms linear in $\beta_{1}$ and $\beta_{2}$.
Then it is natural to introduce the variable substitution
\begin{equation}
    x=\beta_{1}v_{1}+\beta_{2}u_{2}\:,\qquad 
    \tilde{\beta}_{2}=u_{2}\beta_{2}/v_{1}
\end{equation}
so that all letters of $\mathcal{G}_{2}$ are linear in $\tilde{\beta}_{2}$ \footnote{The rescaling $\tilde{\beta}_2=u_2\beta_2/v_1$ is not essential for linear reducibility but is chosen to set the coefficient of $x^4$ in \eqref{ellcurve} to $1$.}. Now the integration over $\tilde{\beta}_{2}$ gives 
\begin{equation} \label{1dintegralofdb}
    I^{\text{ell}}_{\text{db}}=\int_{0}^{\infty} \frac{d  x}{y}\: \mathcal{G}_{3}(x,y)\,,
\end{equation}
where 
\begin{align}
    y^{2}&=x^{4}+a_{3}x^{3}+a_{2}x^{2}+a_{1}x+a_{0} \label{ellcurve} \\
&=\Bigl(\frac{v_{1}}{u_{4}}\bigl((1{-}u_{4})(x{+}1{-}v_{2}){-}u_{1}{+}u_{3}v_{2}\bigr){+}h_{1}{+}h_{2}\Bigr)^{2}{-}4h_{1}h_{2} \nonumber
\end{align}
with
\begin{align}
    h_{1}&=\frac{u_{2}u_{4}}{v_{1}}\bigl(x^{2}+(1{-}u_{1}{+}v_{1})x+v_{1}\bigr),  \\
    h_{2}&=\Bigl(x{+}\frac{v_{1}}{u_{4}}\Bigr)\Bigl((1{+}x{-}u_{1})\Bigl(\frac{u_{2}u_{4}}{v_{1}}-1\Bigr)+(1{-}u_{3})v_{2}\Bigr). \nonumber
\end{align}
Here, the coefficients $a_{i}$ are polynomials in the cross-ratios, and $\mathcal{G}_{3}$ is a pure combination of MPLs of weight three whose letters are rational functions of $x$ and $y$. 

At this stage, there is no obstacle to performing the integration over $x$ and evaluating it in terms of $\mathrm{E}_{4}$ functions which are recursively defined as \cite{Broedel:2017kkb}
\begin{equation}
\label{Eiterateddefinition}
    \Ef{n_1 & \ldots & n_k}{c_1 & \ldots& c_k}{x}=
    \int_{0}^{x}d  x'\,\psi_{n_{1}}(c_{1},x')\Ef{n_{2} & \ldots & n_k}{c_{2} & \ldots& c_k}{x'}
\end{equation}
with $\mathrm{E}_{4}(;x)=1$, where
\begin{equation} \label{psibasis}
    \begin{split}
        &\psi_{0}(0,x)=\frac{1}{y} \,,\qquad  \:\:\psi_{-1}(\infty,x)=\frac{x}{y}\,, \\
        &\psi_{1}(c,x)=\frac{1}{x-c}\,,\quad 
        \psi_{-1}(c,x)=\frac{y_{c}}{y(x-c)}\,,
    \end{split} 
\end{equation}
with $y_{c}=y\vert_{x=c}$. The definition of $\psi_{n}(c,x)$ for $|n|>1$ can be found in \cite{Broedel:2017kkb}; the kernels \eqref{psibasis} are sufficient for the computation of the double-box integral, though. Note that we do not introduce a normalization factor for $\psi_{0}(0,x)$ as in \cite{Broedel:2017kkb}; the reason will be clear in the next section.

We give the final result in terms of $\mathrm{E}_{4}$ functions in the ancillary files. Here, we only record the arguments $c_{i}$ of the $\mathrm{E}_{4}$'s, which make up the set
\begin{align} \label{arguset}
     \biggl\{&0,-1,\infty,-u_{2},-v_{1},-\frac{v_{1}}{u_{4}},-1+\frac{u_{1}}{u_{3}},-u_{2}u_{4}u_{5}, \nonumber \\
    &{-}u_{2}(1{-}u_{4}){-}v_{1},\frac{u_{2}(u_{3}{+}u_{4}{-}1){-}v_{1}}{1-u_{3}},\frac{u_{2}u_{3}u_{4}u_{5}{-}v_{1}}{1-u_{3}}, \nonumber \\
    &\frac{u_{2}u_{3}u_{4}u_{5}{-}v_{1}}{u_{4}(1-u_{3}u_{5})},\frac{u_{2}(u_{3}u_{4}u_{5}v_{2}{-}u_{1})}{u_{3}v_{2}-u_{1}},\frac{v_{1}(u_{3}u_{4}u_{5}v_{2}{-}u_{1})}{u_{4}(u_{1}{-}u_{3}u_{5}v_{2})}, \nonumber \\
    &\frac{u_{4}u_{5}(u_{2}(u_{4}{-}1){-}v_{1}){+}v_{1}}{u_{4}(u_{5}-1)},\frac{u_{1}u_{2}(u_{4}{-}1){-}v_{1}(u_{1}{-}u_{3}v_{2})}{u_{1}-u_{3}v_{2}}, \nonumber \\
    &z_{1,3,5,8}{-}1,\bar{z}_{1,3,5,8}{-}1,z_{1,3,6,8}{-}1,\bar{z}_{1,3,6,8}{-}1,\nonumber \\
    &{-}z_{3,5,8,10},{-}\bar{z}_{3,5,8,10},{-}z_{3,6,8,10},{-}\bar{z}_{3,6,8,10},\nonumber \\
    &\frac{u_{2}u_{3}u_{4}u_{5}{-}v_{1}+ r_{+}}{1-u_{3}} ,\frac{u_{2}u_{3}u_{4}u_{5}{-}v_{1}+ r_{-}}{1-u_{3}} 
    \biggr\}\,,
\end{align} 
where $z_{abcd}\bar{z}_{abcd}=x_{ab;cd}$, $(1-z_{abcd})(1-\bar{z}_{abcd})=x_{da;bc}$, and
\begin{equation}
    r_{\pm} = \frac{\mathcal{G}^{-1}_{45}\operatorname{det}\mathcal{G}\pm\sqrt{\operatorname{det}\mathcal{G}^{(45)}}\sqrt{-\operatorname{det}\mathcal{G}}}{2(1-u_{5})x_{15}^{2}x_{310}^{2}x_{16}^{2}x_{38}^{2}x_{510}^{2}} \:.
\end{equation}
Here, we have introduced the Gram matrix $\mathcal{G}=(x_{ij}^{2})$ with $i$ and $j$ running over the set $\{1,3,5,6,8,10\}$, the elements of the inverse of the Gram matrix $\mathcal{G}^{-1}_{ij}=(\mathcal{G}^{-1})_{ij}$, as well as the matrix $\mathcal{G}^{(ij)}$ obtained from $\mathcal{G}$ by deleting the $i$'th and $j$'th rows and columns.

 \section{From the elliptic curve to the torus: A birational approach}
 \label{sec: to the torus}

 To define the pureness of the double-box integral and to evaluate its symbol, one needs to express it in terms of iterated integrals on the torus. To this end, we need to find a bijection between the elliptic curve $\mathcal{C}$ and the torus $\mathbb{C}/\Lambda$, where $\Lambda$ is the lattice generated by the periods $\omega_{1}$ and $\omega_{2}$ of the elliptic curve. Instead of using the map provided in \cite{Broedel:2017kkb}, here we adopt another strategy: first we find the standard Weierstrass form $Y^{2}=4X^{3}-g_{2}X-g_{3}$ birationally equivalent to $\mathcal{C}$ based on its rational point at infinity \footnote{For a general elliptic curve $y^{2}=\sum_{i=0}^{4} a_{i}x^{i}$ with some rational point $(p,q)$, one can use the variable substitution $(x,y)=(p+\tilde{{x}}^{-1},q \tilde{y}/\tilde{x}^{2})$ to obtain a birational equivalent elliptic curve of form \eqref{ellcurve}.}, then use the standard map $z\mapsto (X,Y)=(\wp(z),\wp'(z))$ in terms of the Weierstrass $\wp$ function. This gives $z\mapsto (x,y)=(\kappa(z),\kappa'(z))$, where
 \begin{equation}
     \kappa(z)=\frac{6a_{1}-a_{2}a_{3}+12 a_{3}\wp(z)-24\wp'(z)}{3a_{3}^{2}-8(a_{2}+6\wp(z))}.
 \end{equation} 
 There are several comments in order: i) there is no need to introduce a normalization factor for $\psi_{0}(0,x)$ as in \cite{Broedel:2017kkb} since $y=\kappa'(z)$, ii) the infinity point $(+\infty,+\infty)$ is mapped to a lattice point.
 iii) each point $c$ in kinematic space corresponds to two points $(c,\pm y_{c})$ on the elliptic curve $\mathcal{C}$ and hence to two images on the torus $\mathbb{C}/\Lambda$, which we denote by $z_{c}^{\pm}$;
 these two images satisfy 
 \begin{equation} \label{wpwmrelation}
     z_{c}^{+}+z_{c}^{-}= z^{-}_{\infty} + z^{+}_{\infty}\equiv z^{-}_{\infty} \:  \operatorname{mod} \Lambda \:,
 \end{equation}
 since the corresponding points $(X_{c}^{\pm},Y_{c}^{\pm})$, together with $(X_{\infty}^{-},Y_{\infty}^{-})$,
 are on the same line. Similarly, one can find that the torus images $z_{c_{i}}^{\pm}$ of the kinematics $c_{i}$ in  \eqref{arguset} satisfy 
 \begin{align}
     &\{z^{+}_{c_{6}}+z^{+}_{c_{7}},z_{c_{25}}^{+}+z^{+}_{c_{26}},z^{+}_{c_{6}}+z^{+}_{c_{10}}+z^{+}_{c_{16}},  \\
      &\quad z^{+}_{c_{10}}-z^{+}_{c_{17}}+z^{+}_{c_{18}},z^{+}_{c_{16}}+z^{+}_{c_{23}}-z^{+}_{c_{24}}\}\equiv 0 \:\operatorname{mod}\Lambda\:, \nonumber
 \end{align}
 and similar relations for $z^{-}_{c_{i}}$ obtained by  \eqref{wpwmrelation}, where $c_{i}$ is the $i$'th element in  \eqref{arguset}.

 \tikzset{cross/.style={cross out, thick, draw=black, fill=none, minimum size=2*(#1-\pgflinewidth), inner sep=0pt, outer sep=0pt}, cross/.default={2pt}}

 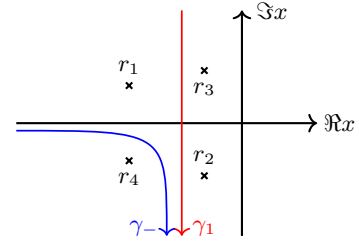
\begin{figure}
    \centering
    \begin{tikzpicture}
            \draw[->, thick] (0,0) to (0,3);
            \draw[->, thick] (-3,1.5) to (1,1.5);
            \draw[<-,line width=0.23mm,red] (-0.8,0.0) to (-0.8,3.0);
            \draw[->,line width=0.23mm,blue] (-3,1.4) .. controls (-1,1.4) ..   (-1.0,0.0) ; 
            \node[cross,label=above:$r_{1}$] at (-1.5, 2.0) {};
            \node[cross,label=below:$r_{4}$] at (-1.5, 1.0) {};
            \node[cross,label=above:$r_{2}$] at (-0.5, 0.8) {};
            \node[cross,label=below:$r_{3}$] at (-0.5, 2.2) {};
            \node at (-0.5,0.1) {\textcolor{red}{$\gamma_{1}$}};
            \node at (-1.3,0.1) {\textcolor{blue}{$\gamma_{-}$}};
            \node at (0.4,3) {$\Im x$};
            \node at (1.3,1.5) {$\Re x$};
    \end{tikzpicture}
    \caption{Four roots of $y^{2}(x)$ in the positive kinematics region and two integration contours. The contour $\gamma_2$ which defines $\omega_2$ runs along the real axis.
    }
    \label{fig: contours}
    \end{figure}

 In what follows, we will work in the region given by positive momentum-twistor kinematics 
 \cite{Hodges:2009hk,Arkani-Hamed:2013jha}, where the four roots of $y^{2}(x)$ come in complex conjugate pairs as shown in fig.\ \ref{fig: contours}. In this case, the torus image $z_{c}^{+}$ for any real $c$ is simply given by
\begin{equation}
    z_{c}^{+}=\int_{-\infty}^{c} \frac{dx}{y}\,.
\end{equation} 
Hence, $z_{\infty}^{+}$ is one period of the torus, and we choose it to be $\omega_{2}$. The image $z_{c}^{-}$ can be obtained by \eqref{wpwmrelation} together with $z_{\infty}^{-}=\int_{\gamma_{-}} dx/y$, and the other period is $\omega_{1}=\int_{\gamma_{1}} dx/y$, where the integration contours are defined in fig.\ \ref{fig: contours}.

Now one can introduce iterated integrals on the normalized torus with periods $(1,\tau=\omega_{2}/\omega_{1})$ \cite{Broedel:2017kkb,Broedel:2018iwv}:
\begin{equation}
    \gamt{n_1 & \ldots & n_k}{w_1 & \ldots& w_k}{w}=
    \int_{0}^{w}d w'\,g^{(n_{1})}(w'{-}w_{1})\gamt{n_{2} & \ldots & n_k}{w_{2} & \ldots& w_k}{w'}
\end{equation}
with $\tilde{\Gamma}(;w)=1$. Such an iterated integral is said to have length $k$ and weight $\sum_kn_k$, and in contrast to the case of MPLs, the two are not necessarily equal. The integration kernels $g^{(n)}(z)$ are generated by the \emph{Eisenstein-Kronecker series}
\begin{equation}
    \frac{\partial_{z}\theta_{1}(0)\theta_{1}(z+\alpha)}{\theta_{1}(z)\theta_{1}(\alpha)} = \sum_{n\geq 0}\alpha^{n-1}g^{(n)}(z)\:,
\end{equation}
where $\theta_{1}(z)=\theta_{1}(z|\tau)$ is the odd Jacobi theta function.

With these conventions, it is not hard to find
\begin{subequations} \label{psitokg}
\begin{align}
    \psi_{1}(c,x)d  x&=\bigl[g^{(1)}(w-w_{c}^{+})+g^{(1)}(w-w_{c}^{-}) \label{psitokga}  \\
    &\qquad - g^{(1)}(w-w_{\infty}^{+})-g^{(1)}(w-w_{\infty}^{-}) \bigr] dw \,,  \nonumber  \\
    \psi_{-1}(c,x)d  x&=\bigl[g^{(1)}(w-w_{c}^{+})-g^{(1)}(w-w_{c}^{-})  \label{psitokgb} \\
    &\qquad + g^{(1)}(w_{c}^{+})-g^{(1)}(w_{c}^{-}) \bigr] dw\,,   \nonumber  \\
    \psi_{-1}(\infty,x) dx&=\bigl[g^{(1)}(w-w_{\infty}^{-})-g^{(1)}(w)    \label{psitokgc}\\
    &\qquad +g^{(1)}(w_{\infty}^{-})-\omega_{1}a_{3}/4 \bigr]dw \,, \nonumber
\end{align}
\end{subequations}
as well as $\psi_{0}dx= \omega_{1}dw$, where $w_{c}^{\pm}$ are the normalized torus images $z_{c}^{\pm}/\omega_{1}$. It is then trivial to express the double-box integral in terms of $\tilde{\Gamma}$-functions as \footnote{Without expressing the double-box in terms of eMPLs, it was previously argued that it is pure in \cite{Broedel:2018qkq}.}
\[
    I_{\text{db}}^{\text{ell}} = \omega_{1} T_{\text{db}}^{\text{ell}} \:,
    \]  
where $T_{\text{db}}^{\text{ell}}$ is a pure combination of $\tilde{\Gamma}$'s of length four and weight three.

Equivalently, the functions $\tilde{\Gamma}$ can be expressed in terms of the functions $\mathcal{E}_4$ \cite{Broedel:2018qkq}, which are defined in complete analogy to \eqref{Eiterateddefinition} in terms of kernels ($n\geq0$)
\begin{align}
 \Psi_{\pm n}(c,x)d  x&=\bigl[g^{(n)}(w-w_{c}^{+})\pm g^{(n)}(w-w_{c}^{-}) \label{Psikernels}  \\
    &- \delta_{\pm n,1}(g^{(1)}(w-w_{\infty}^{+})+g^{(1)}(w-w_{\infty}^{-})) \bigr] dw\,.\nonumber
\end{align}
We provide the more compact expression for $T_{\text{db}}^{\text{ell}}$ in terms of $\mathcal{E}_4$'s, as well as code expanding it in terms of $\tilde{\Gamma}$'s, in the ancillary files. 

Let us close this section by remarking on the shuffle regularization. It will be convenient to introduce $\Omega^{(j)}$, defined via 
\begin{equation} \label{letterdef}
\begin{aligned}
 \partial_z\Omega^{(j)}(z,\tau)&=(2\pi i)^{1-j}g^{(j)}(z,\tau)\,,\\
  \partial_\tau\Omega^{(j)}(z,\tau)&=j(2\pi i)^{-j}g^{(j+1)}(z,\tau)\,.
  \end{aligned}
\end{equation}
As we shall see in the next section, these $\Omega^{(j)}$ appear as the symbol letters. In contrast to \cite{brown2011multiple,Broedel:2018iwv}, we have included factors of $2\pi i$ such that all letters have weight 1 and we can find linear relations with rational coefficients among them. By definition, $(2\pi i)^{1-j} \gamt{j}{0}{w}= \Omega^{(j)}(w)- \Omega^{(j)}(0)$. However, $\Omega^{(1)}(w)$ is singular at $w=0$, and the usual shuffle regularization \cite{Broedel:2018iwv} takes $\gamt{1}{0}{w}=\Omega^{(1)}(w)-2\log\eta(\tau)$ with Dedekind eta function $\eta(\tau)$. Here, to be consistent with the shuffle regularization $G(0;x)=\log(x)$ for MPLs which we implicitly used in \eqref{1dintegralofdb}, we take the shuffle regularization of $\gamt{1}{0}{w}$ to be
\begin{equation}
    \gamt{1}{0}{w}= \Omega^{(1)}(w)-2\log\eta(\tau) - \log \frac{2\pi  i }{\omega_{1}y_{0}} \:.
\end{equation}

\section{Symbology}
\label{sec: symbol}

The symbol of $\widetilde{\Gamma}$ can be defined recursively via the differential of  $\widetilde{\Gamma}$ in a similar way as for MPLs \cite{Broedel:2018iwv}.
The differential of $\widetilde{\Gamma}_{k}^{(n)}$ of weight $n$ and length $k$ schematically takes the form 
\begin{equation}
 d\widetilde{\Gamma}_{k}^{(n)}=\sum_i  (2\pi i)^{j_i-1} \tilde{\Gamma}^{(n-j_{i})}_{k-1}
d\Omega^{(j_i)}(y_i)
 \,,
\end{equation}
where the $\Omega^{(j)}$ are given in \eqref{letterdef} with $\Omega^{(-1)}=-2\pi i \tau$; the precise formula is given in \cite{Broedel:2018iwv}. It is easy to see that there would be an overall factor $(2\pi i)^{n-k}$ if we keep taking the differential recursively. Thus, it is natural to define the symbol for $(2\pi i)^{k-n}\tilde{\Gamma}_{k}^{(n)}$ rather than $\tilde{\Gamma}_{k}^{(n)}$ as 
\begin{equation}
 \mathcal{S}((2\pi i)^{k-n}\widetilde{\Gamma}_{k}^{(n)})=\sum_i \mathcal{S}((2\pi i)^{k-n+j_{i}-1}\tilde{\Gamma}_{k-1}^{(n-j_{i})})\otimes \Omega^{(j_i)}
 \,.
\end{equation}

For the double box, the resulting symbol is of the form
\begin{equation}
    \mathcal{S}(T_{\text{db}}^{\text{ell}})=\frac{1}{2\pi i} \sum_{j}\Omega^{(j_{1})}(w_{j_{1}})\otimes\cdots \otimes\Omega^{(j_{4})}(w_{j_{4}}),
\end{equation}
where $\sum_{i=1}^{4}j_{i}=3$. Naively, there would be $\Omega^{(6)}$'s at most due to the existence of $\Omega^{(-1)}$, but all $\Omega^{(j>3)}$'s drop out after using $\Omega^{(j)}(-w)=(-1)^{j+1}\Omega^{(j)}(w)$. At this stage, the symbol has around $10^{6}$ terms. However, it is worth noting that -- due to including factors of $2\pi i$ in the definition of $\Omega^{(j)}$ -- all these letters have weight 1 despite their superscripts and further simplifications hence are possible.

To make contact with the more familiar kinematic world, we can apply \eqref{psitokga} to $\int_{a}^{b} \psi_{1}(c,x)dx$ to derive the following identity:
\begin{align} \label{identityproved}
    &\quad \log\frac{c-a}{c-b} +\sum_{\sigma\in \pm} \Omega^{(1)}(w_{c}^{\sigma}-w_{b}^{+})-\Omega^{(1)}(w_{c}^{\sigma}-w_{a}^{+}) \nonumber \\
    &= \sum_{\sigma\in \pm}\Omega^{(1)}(w_{\infty}^{\sigma}-w_{b}^{+})-\Omega^{(1)}(w_{\infty}^{\sigma}-w_{a}^{+})\,.
\end{align}
Further identities involving elliptic letters can be found using the PSLQ algorithm after numerically evaluating the letters via the sum representations given in the supplementary material. 
For example, we found
\begin{align}
    &\sum_{i=1}^{6}(-1)^{i+1}\Bigl(\Omega^{(1)}(w_{d_{i}}^{-}-w_{\infty}^{+})-\Omega^{(1)}(w_{d_{i}}^{-}-w_{0}^{+})\Bigr)  \nonumber \\
    &\equiv\log \frac{d_{2}}{d_{3}d_{5}}+\Omega^{(0)}(w_{\infty}^{+}-w_{0}^{+}) \: \operatorname{mod} i\pi
\end{align}
with 
$
    d_{i} \in \{\infty, {-}v_{1}/u_{4}, \bar{z}_{1,3,5,8}-1, z_{1,3,5,8}-1, 
   -z_{3,6,8,10},$ $ -\bar{z}_{3,6,8,10}\}
   $.
Moreover, we found complicated identities involving $\Omega^{(2)}$. 
All these identities turn out to be consequences of Abel's addition theorem and the elliptic Bloch relation~\cite{Zagier2000,bloch2011higher,Broedel:2019tlz}.  

Combining \eqref{identityproved} and the identities found via the PSLQ algorithm,
a dramatic simplification happens: all $\Omega^{(3)}$'s drop out, only logs remain in the first two entries, and the symbol ends up with an expression of around $10^{4}$ terms!
Remarkably, the resulting simplified symbol satisfies the same physical first entry conditions found in the MPL case \cite{Gaiotto:2011dt}, that is the first entries can only be $\log(x_{ab;cd})$. Moreover, the symbol follows certain patterns for the first two entries observed in the MPL case in \cite{Gaiotto:2011dt,CaronHuot:2011ky,Dennen:2015bet,He:2020lcu}: the first two entries form the symbols of $\operatorname{Li}_{2}(1-x_{ab;cd})$, $\log(x_{ab;cd})\log(x_{a'b';c'd'})$ or four-mass boxes. In particular, the symbol satisfies the Steinmann conditions \cite{Steinmann,Steinmann2}, i.e.\ discontinuities in partially overlapping channels vanish.

The complete symbol can be organized by its seven elliptic last entries of type $\Omega^{(0)}(w,\tau)=2\pi i w$ as well as by its behaviour under the two reflections $R_1$, $R_2$:
\begin{align} \nonumber
    & \mathcal{S}(T_{\text{db}}^{\text{ell}}) = \mathcal{S}(I_\text{hex}) \otimes( w^{+}_{c_{25}}{-}\tfrac{w^{+}_{\infty}}{2})
    +\mathcal{S}(F_{-})\otimes (w_{\infty}^{-}{-}\tfrac{w_{\infty}^{+}}{2})  \\
    &+\mathcal{S}(F_{+})\otimes w_{\infty}^{+}+ \Bigl[\mathcal{S}(F_{17})\otimes (w_{c_{17}}^{+}{-}\tfrac{w_{\infty}^{+}}{2})+ \text{reflections}\Bigr] , \label{diffofdb}
\end{align}
where $I_{\text{hex}}$ is the 6D hexagon integral (normalized to be pure) in fig.\ \ref{fig: integral} and $F_+$, $F_-$, $F_{17}$ are weight-3 functions whose symbols are known from $\mathcal{S}(T_{\text{db}}^{\text{ell}})$ and recorded in an ancillary file.
In particular, $I_{\text{hex}}$, $F_-$, and $F_{17}$ are polylogarithmic. 
The symbol can be written in terms of 36 rational letters, 24 algebraic letters (in terms of momentum twistors \cite{Hodges:2009hk}), and besides the 7 elliptic last entries, elliptic letters only appear at the third entry of $\mathcal{S}(F_{+})$ and come in only 13 linear independent combinations! (For a list of symbol letters, see the supplementary material and the ancillary files.)
The first three terms in \eqref{diffofdb} are individually invariant under $R_{1}$ and $R_2$.
The fourth term generates a 4-orbit, as indicated by the ``$+$ reflections''.
The precise behaviour of the torus images under the reflections is given in the supplementary material.

Finally, let us remark on the differential equation relating the double box to the 6D hexagon \cite{Paulos:2012nu,Nandan:2013ip}. At the level of the symbol it becomes an immediate consequence of \eqref{diffofdb} since only $w^{+}_{c_{25}}$ in the 7 last entries depends on $u_5$ and $ \omega_{1}\partial_{u_{5}} w^{+}_{c_{25}}=x_{16}^{2}x^{2}_{38}x_{510}^{2}/\sqrt{-\operatorname{det}\mathcal{G}}$.

\section{Conclusion and Outlook}
\label{sec: conclusion}

In this letter, we have calculated the 10-point two-loop massless double-box integral in terms of eMPLs and calculated its symbol. This integral is the sole contribution to a particular component of the 10-point N${}^3$MHV superamplitude in planar $\mathcal{N}=4$ SYM theory, thus allowing us to draw direct conclusions from our findings for scattering amplitudes.

We find that the symbol of the double-box integral shows a very rich structure. In particular, the first entry of the symbol is drawn from the letters $\log(x_{ab;cd})$, where $x_{ab;cd}$ is a dual-conformal cross-ratio. This means that the double-box integral, despite being elliptic, satisfies exactly the same first-entry conditions that were argued to occur for amplitudes built from non-elliptic polylogarithms. 
Moreover, the second entry of the symbol contains only letters of $\log$ type and satisfies patterns previously observed in the non-elliptic case. 
The last entry of the symbol is also very restricted, containing only seven possible letters, of elliptic type $\Omega^{(0)}$.

Taking the symbol of our result for the double-box integral, we observed massive cancellations and simplifications, partially due to identities which we first observed numerically via the PSLQ algorithm. 
As we will elaborate in upcoming work \cite{in_progress}, these identities are consequences of Abel's addition theorem and the elliptic Bloch relation~\cite{Zagier2000,bloch2011higher,Broedel:2019tlz}.
It would be interesting to use similar identities to better understand the 13 linearly independent combinations in which the elliptic letters occur in the third entry. 
Moreover, it would be very interesting to lift this simplified symbol to a simplified function.

The symbol of the double-box integral manifests the differential equation relating it to the 6D one-loop hexagon integral. 
This suggests that one can bootstrap the symbol via this differential equation, i.e.\ taking the known symbol of the hexagon, appending the elliptic final letter corresponding to the differential equation and constructing the remainder of the symbol by imposing integrability \cite{in_progress}. Schematically,
\begin{equation}
 \mathcal{S}\Bigl(\!\!\begin{aligned}
 \begin{tikzpicture}[scale=0.25]
		\node (0) at (0.5, 15.5) {};
		\node (1) at (2, 16) {};
		\node (2) at (1, 16) {};
		\node (3) at (3, 16) {};
		\node (4) at (3.5, 15.5) {};
		\node (5) at (1, 14) {};
		\node (6) at (0.5, 14.5) {};
		\node (7) at (2, 14) {};
		\node (8) at (3.5, 14.5) {};
		\node (9) at (3, 14) {};
		\draw[thick] (0.center) to (4.center);
		\draw[thick] (6.center) to (8.center);
		\draw[thick] (2.center) to (5.center);
		\draw[thick] (1.center) to (7.center);
		\draw[thick] (3.center) to (9.center);
\end{tikzpicture} 
\end{aligned}
\!\!\Bigr)
=\mathcal{S}\Bigl(\begin{aligned}
 \begin{tikzpicture}[scale=0.25]
       \draw [thick] (4.9,15.4) -- (4.9,14.6);
       \draw [thick] (6.2,15.4) -- (6.2,14.6);
       \draw [thick] (6.2,15.4) -- (5.55,15.8) -- (4.9,15.4);
       \draw [thick] (4.9,14.6)-- (5.55,14.2) --  (6.2,14.6);
       \draw [thick] (5.55,15.8) --(5.55,16.1) ;
       \draw [thick] (5.55,14.2)-- (5.55,13.9);
       \draw [thick] (4.9,15.4) -- (4.6, 15.7);
       \draw [thick] (4.9,15.4) -- (4.5, 15.5);
       \draw [thick] (4.9,14.6) -- (4.6, 14.3);
       \draw [thick] (4.9,14.6) -- (4.5, 14.5); 
       \draw [thick] (6.2,15.4) -- (6.5,15.7);
       \draw [thick] (6.2,15.4) -- (6.6,15.5);
       \draw [thick] (6.2,14.6) -- (6.5,14.3);
       \draw [thick] (6.2,14.6) -- (6.6,14.5); 
\end{tikzpicture} 
\end{aligned}
\Bigr)
\otimes w_{c_{25}}^{+}+\text{integrability}.
\end{equation}

Traintrack integrals \cite{Bourjaily:2018ycu}, which involve integrations over higher-dimensional Calabi-Yau manifold, similarly satisfy differential equations relating them to $n$-gons \cite{Paulos:2012nu,
Nandan:2013ip}.
While the functional space and corresponding symbol is not yet understood in these cases, it seems likely that a bootstrap based on the differential equation will also be possible in these cases.

In the case of MPLs with rational arguments, the symbol alphabets occurring for amplitudes as well as their adjacency conditions can be understood in terms of cluster algebras \cite{Drummond:2017ssj,Drummond:2018dfd,Drummond:2018caf}, and a similar understanding is currently being developed in the case of the Feynman integrals \cite{Chicherin:2020umh,He:2021esx,He:2021non} and amplitudes including algebraic letters \cite{Drummond:2019cxm,Arkani-Hamed:2019rds,Henke:2019hve,Herderschee:2021dez,Henke:2021avn,Ren:2021ztg}. It would be very interesting to use the data we provide in this work to extend the cluster program to the elliptic case.
Similarly, it would be interesting to extend the amplitude bootstrap program to the elliptic case.

The double-box integral we considered in this letter is arguably the simplest elliptic integral contributing to planar $\mathcal{N}=4$ SYM theory, and to massless QFTs in general. We expect that the techniques developed in this letter can also be applied to more general elliptic integrals, such as the general
12-point double-box integral, corresponding penta-box integrals and double-pentagon integrals. 
Combining these integrals with the understanding of prescriptive unitarity and the corresponding leading singularities \cite{Bourjaily:2020hjv,Bourjaily:2021vyj} would directly allow to calculate many further elliptic amplitudes in the massless case.
Moreover, also the double-box integral with generic masses is elliptic \cite{Bloch:2021hzs}, and should be amenable to the techniques presented here.

\begin{acknowledgments}
We thank Andrew McLeod and Mark Spradlin for comments on the manuscript, Claude Duhr and Robin Marzucca for communication as well as Cristian Vergu and Matthias Volk for discussions.
CZ is grateful to Zhenjie Li for his instruction on some issues of numeric computations.
This work was supported by the research grant  00025445 from Villum Fonden and the ERC starting grant 757978.

\end{acknowledgments}

\bibliography{reference} 

\widetext

\begin{center}
\textbf{\large Supplemental materials}
\end{center}
\setcounter{equation}{0}
\setcounter{figure}{0}
\setcounter{table}{0}
\makeatletter
\renewcommand{\theequation}{A\arabic{equation}}
\renewcommand{\thefigure}{A\arabic{figure}}
\renewcommand{\bibnumfmt}[1]{[A#1]}
\renewcommand{\citenumfont}[1]{A#1}

\section*{Several useful sum expressions}

Here we record useful sum expressions for the integration kernels $g^{(j)}(z)$ and the elliptic letters $\Omega^{(j)}$, which are valid for $j$ odd/even:
\begin{align} \label{gdef}
    g^{(\text{odd } j)}(z)&= -\pi i\biggl( \frac{1+e^{2\pi i z}}{1-e^{2\pi i z}}\biggr) \delta_{j,1} 
    +\frac{(2\pi i)^{j}}{(j{-}1)!}\sum_{n=1}^{\infty}
    n^{j-1}\biggl(\frac{1}{1-\me^{2\pi i (n\tau-z)}}- \frac{1}{1-\me^{2\pi i (n\tau+z)}}\biggr), \nonumber \\ 
    g^{(\text{even } j)}(z)&=-2\zeta_{j}
    - \frac{(2\pi i)^{j}}{(j{-}1)!}\sum_{n=1}^{\infty} n^{j-1}
    \biggl(\frac{e^{2\pi i (n\tau+z)}}{1-e^{2\pi i (n\tau+z)}}+\frac{e^{2\pi i (n\tau-z)}}{1-e^{2\pi i (n\tau-z)}}\biggr) ,
\end{align}
and
\begin{align} \label{wdef}
    \Omega^{(\text{odd } j)}(z)&=\Bigl(\log(1-\me^{2\pi i  z})-\pi i  z\Bigr) \delta_{j,1}- \frac{2j\zeta_{j+1}\tau}{(2\pi i)^{j}} + 
    \frac{1}{(j{-}1)!}\sum_{n=1}^{\infty}
    n^{j-1}\log\bigl((1-\me^{2\pi i (n\tau-z)})(1-\me^{2\pi i (n\tau+z)})\bigr), \nonumber \\ 
    \Omega^{(\text{even } j)}(z)&=-\frac{2\zeta_{j}z}{(2\pi i)^{j-1}} 
    + \frac{1}{(j{-}1)!}\sum_{n=1}^{\infty} n^{j-1}\log\frac{1-\me^{2\pi i (n\tau+z)}}{1-\me^{2\pi i (n\tau-z)}} ,
\end{align}
where $\zeta_{j}=\sum_{n\in\mathbb{Z}_{+}} n^{-j}$ are the usual zeta values.

\section*{Reflection symmetries of the double box}

The reflections $R_1$ and $R_2$ act on the dual coordinates as
\begin{equation}
 \{x_{5},x_{3},x_{1},x_{10},x_{8},x_{6}\}\xleftarrow{R_1}\{x_{1},x_{3},x_{5},x_{6},x_{8},x_{10}\}\xrightarrow{R_2}\{x_{10},x_{8},x_{6},x_{5},x_{3},x_{1}\}\,.
\end{equation}
Consequently, they act on the cross-ratios as
\begin{equation}
 \{v_{1},u_{1},v_{2},u_{2},u_{4},u_{3},u_{5}\}\xleftarrow{R_1}\{u_{1},v_{1},u_{2},v_{2},u_{3},u_{4},u_{5}\}\xrightarrow{R_2}\{u_{2},v_{2},u_{1},v_{1},u_{4}u_{2}/v_{1},u_{3}v_{2}/u_{1},u_{5}\}\,.
\end{equation}
To be consistent with the action on momentum twistors, the action of $R_{2}$ on $\sqrt{-\mathcal{G}}$ gives $-\sqrt{-\mathcal{G}}$. 

The actions on the torus images are less trivial, but one can show that $w_{\infty}^{+}$ is invariant under both reflections, $w_{c_{25}}^{+}$ is invariant under the action of $R_{1}$, while $R_{2}: w_{c_{25}}^{+}-(w_{\infty}^{+}/2)\mapsto(w_{\infty}^{+}/2)-w_{c_{25}}^{+}$ and $R_{1,2}: w_{\infty}^{-}-(w_{\infty}^{+}/2)\mapsto(w_{\infty}^{+}/2)-w_{\infty}^{-}$ (up to an integer real part).
Moreover, 
\begin{equation}
    \begin{tikzcd}
        w_{c_{17}}^{+}{-}(w_{\infty}^{+}/2) \arrow[r, "R_{1}"] \arrow[d,"R_{2}"]
        &(w_{\infty}^{+}/2) - w_{c_{18}}^{+} \arrow[d, "R_{2}" ] \\
        (w_{\infty}^{+}/2) - w_{c_{23}}^{+} \arrow[r,  "R_{1}" ]
        &   w_{c_{24}}^{+}{-}(w_{\infty}^{+}/2)
    \end{tikzcd} \:.
\end{equation}

\section*{Symbol Alphabet}

The symbol can be written in terms of 36 rational letters, 24 algebraic letters, 7 simple elliptic letter and 13 complicated linear combinations of elliptic letters. 
Note that the symbol alphabet can potentially be further reduced based on an improved understanding of the 13 linear combinations of elliptic letters. 

The rational and algebraic letters can be conveniently expressed in terms of momentum twistors $Z_{i}=(\lambda_{i}^{\alpha},x_{i}^{\alpha\dot{\alpha}}\lambda_{i\alpha})$ with the dual coordinate $x_{i}$ and spinor-helicity variable $\lambda_{i}$, for which we use the following standard notations: $\langle ijkl\rangle= \epsilon_{abcd}Z_{i}^{a}Z_{j}^{b}Z_{k}^{c}Z_{l}^{d}$, $\langle a(bc)(de)(fg)\rangle =\langle abde\rangle \langle acfg \rangle-\langle acde\rangle \langle abfg \rangle$, 
$\langle ab(cde)\cap(ijk)\rangle=\langle acde\rangle\langle bijk\rangle-\langle bcde\rangle\langle aijk\rangle$ as well as 
$(ab)\cap(ijk)=Z_{a}\langle bijk\rangle+Z_{b}\langle ijka\rangle$.

We find the following sets of letters.

\noindent 1. Rational letters:
    \begin{itemize}
        \item $\langle 1,2,3,10\rangle$, $\langle 1,4,5,10\rangle$, $\langle 1,5,6,10\rangle$, 
        $\langle 1,7,8,10 \rangle$, $\langle 1,5,9,10\rangle$, $\langle 2,3,5,10\rangle$ and 10 others generated by the two reflections.
        \item $\langle 5\, (4,6)(7,8)(9,10)\rangle$, $\langle 5 \,(2,3)(7,8)(9,10)\rangle$, $\langle 5 \,(2,3)(4,6)(9,10)\rangle$, $\langle 5\, (2,3)(4,6)(7,8)\rangle$ and 10 others generated by the two reflections.
        \item $\langle 2,3\, (4,5,6)\cap (9,10,1)\rangle$ and $\langle 7,8 (4,5,6)\cap (9,10,1)\rangle$.
        \item $\langle (2,3)\cap(4,5,6)\,7,8\,(9,10)\cap(4,5,6)\rangle$ and 3 others generated by the two reflections.
    \end{itemize}
    2. Algebraic letters (note that only 24 of these algebraic letters are multiplicatively independent):
    \begin{itemize}
        \item 
$\displaystyle
        \frac{\bar{z}_{1,3,5,8}}{z_{1,3,5,8}}, \frac{1-z_{1,3,5,8}}{1-\bar{z}_{1,3,5,8}}, 
        \frac{\frac{\langle 5 \,(2,3)(4,6)(7,8)\rangle \langle 10,1,2,3\rangle}{\langle 5 \,(2,3)(4,6)(10,1)\rangle \langle 2,3,7,8\rangle}-z_{1,3,5,8}}{\frac{\langle 5 \,(2,3)(4,6)(7,8)\rangle \langle 10,1,2,3\rangle}{\langle 5 \,(2,3)(4,6)(10,1)\rangle \langle 2,3,7,8\rangle}-\bar{z}_{1,3,5,8}} , 
        \frac{\frac{\langle 10 (2,3)(7,8)(9,1)\rangle \langle 4,5,7,8\rangle }{\langle 10 (4,5)(7,8)(9,1)\rangle \langle 2,3,7,8\rangle} -z_{1,3,5,8}}{\frac{\langle 10 (2,3)(7,8)(9,1)\rangle \langle 4,5,7,8\rangle }{\langle 10 (4,5)(7,8)(9,1)\rangle \langle 2,3,7,8\rangle} -\bar{z}_{1,3,5,8}},  
$
$ \displaystyle
        \frac{\frac{\langle 7,8 (4,5,6)\cap(10,2,3)\rangle}{\langle 2,3,7,8\rangle\langle 4,5,6,10\rangle}-z_{1,3,5,8} }{\frac{\langle 7,8 (4,5,6)\cap(10,2,3)\rangle}{\langle 2,3,7,8\rangle\langle 4,5,6,10\rangle}-\bar{z}_{1,3,5,8}}$  and 5 others generated by the reflection $R_{2}$.
        \item 
        $\displaystyle\frac{\bar{z}_{1,3,6,8}}{z_{1,3,6,8}}, \frac{1-z_{1,3,6,8}}{1-\bar{z}_{1,3,6,8}}, 
         \frac{\frac{\langle 7,8 (4,5,6)\cap(10,2,3)\rangle}{\langle 2,3,7,8\rangle\langle 4,5,6,10\rangle}-z_{1,3,6,8} }{\frac{\langle 7,8 (4,5,6)\cap(10,2,3)\rangle}{\langle 2,3,7,8\rangle\langle 4,5,6,10\rangle}-\bar{z}_{1,3,6,8}},
         \frac{\frac{\langle 2,3, (9,10,1)\cap(5,7,8)\rangle}{\langle 2,3,7,8\rangle\langle 1,5,9,10\rangle}-z_{1,3,6,8} }{\frac{\langle 2,3, (9,10,1)\cap(5,7,8)\rangle}{\langle 2,3,7,8\rangle\langle 1,5,9,10\rangle}-\bar{z}_{1,3,6,8}}$  and 4 others generated by the reflection $R_{2}$.
        \item 
        $\displaystyle
        \frac{\bar{z}_{1,3,5,8}-z_{1,3,6,8}}{\bar{z}_{1,3,5,8}-\bar{z}_{1,3,6,8}},\frac{z_{1,3,5,8}-z_{1,3,6,8}}{z_{1,3,5,8}-\bar{z}_{1,3,6,8}}$  and 6 others generated by the two reflections.
 
    \end{itemize}
3. Elliptic letters: besides the 7 last entries, they enter the symbol in terms of 13 independent linear combinations which are recorded in the ancillary files.

\end{document}